\documentclass{revtex4}
\usepackage{amsfonts}
\usepackage{amsmath}

\input{tcilatex}

\begin{document}

\preprint{}
\title{$2+1$ dimensional magnetically charged solutions in Einstein - Power
- Maxwell theory.}
\author{S. Habib Mazharimousavi}
\email{habib.mazharimousavi@emu.edu.tr}
\author{O. Gurtug}
\email{ozay.gurtug@emu.edu.tr}
\author{M. Halilsoy}
\email{mustafa.halilsoy@emu.edu.tr}
\author{O. Unver}
\email{ozlem.unver@emu.edu.tr }
\affiliation{Department of Physics, Eastern Mediterranean University, G. Magusa, North
Cyprus, Mersin 10 - Turkey.}

\begin{abstract}
We obtain a class of magnetically charged solutions in $2+1$ dimensional
Einstein - Power - Maxwell theory. In the linear Maxwell limit, such
horizonless solutions are known to exist. We show that in $3D$ geometry,
black hole solutions with magnetic charge does not exist even if it is
sourced by power-Maxwell field. Physical properties of the solution with
particular power $k$ of the Maxwell field is investigated. The true timelike
naked curvature singularity develops when $k>1$ which constitutes one of the
striking effects of the power Maxwell field. For specific power parameter $k$%
, the occurrence of timelike naked singularity is analysed in quantum
mechanical point of view. Quantum test fields obeying the Klein - Gordon and
the Dirac equations are used to probe the singularity. It is shown that the
class of static pure magnetic spacetime in the power Maxwell theory is
quantum mechanically singular when it is probed with fields obeying
Klein-Gordon and Dirac equations in the generic case.
\end{abstract}

\maketitle

\section{INTRODUCTION}

Unlike the case of $4-$dimensional spacetime, gravitational and
electromagnetic fields in $2+1-$ dimensions ($3D$) show significant
differences. The absence of free gravitational field (or Weyl curvature) in $%
3D$ for instance, is one such noteworthy property as far as gravity is
concerned. The addition of extra sources beside the cosmological constant,
therefore, becomes indispensable to turn this reduced dimension into an
attractive arena for doing physics. We recall the Reissner-Nordstr\"{o}m
(RN) example in which there is a symmetric duality between the electric and
magnetic fields. That is, dual of Maxwell field $2-$form in $4-$dimensions
is still a $2-$form. In $3D,$ on the other hand, duality maps a $2-$form
into $1-$form and vice versa. Besides, the interpretation of the sources of
the electric fields in $3D$ is not ambiguous, however, considering the
magnetic sources the interpretation is not much clear. Yet, for a number of
reasons, which can be summarized as - contributing to our understanding of
their $4-$dimensional counterparts - the $3D$ solutions persist to be a
center of attraction in general relativity. The prototype example of such $%
3D $ black hole solutions is known to be the BTZ \cite{1}. This black hole
was sourced by a mass, a static electric field and a negative cosmological
constant. The existence of magnetically charged $3D$ solutions was also
addressed shortly after BTZ \cite{2,3,4,5}. Dias and Lemos have studied
magnetic solutions in $3D$ Einstein theory including the rotating version 
\cite{6} of the works cited in \cite{2,3,4,5} and also the magnetic point
sources in Brans - Dicke theories \cite{7}. The common result verified,
among found solutions, the absence of such magnetic black holes. In other
words, $3D$ Einstein-Maxwell (EM) equations do not admit a solution that can
be interpreted as a black hole with pure magnetic fields. Furthermore, these
solutions are free of curvature singularities. The nonsingular magnetic
Melvin universe \cite{8} in $4-$dimensions is well-known to provide
information about the existence of such solutions in different dimensions as
well. As a matter of fact, a magnetic solution has physically radical
differences in comparison with its electric counterpart which are related by
a duality transformation \cite{5,9,10}. Although pure magnetic black holes
in $3D$ are yet to be found, we may anticipate that they are crucial in
understanding the global entropic flow and storage / loss of information in
such lower dimensions.

In this paper, we wish to go beyond linear Maxwell electromagnetism and to
consider the recently-fashionable nonlinear electrodynamics (NED) coupled
with gravity in the presence of a negative cosmological constant. This
formalism has already found applications \cite{11,12,13,14,15,16}, but to
the best of our knowledge, in $3D$ pure magnetic version of the power-law,
nonlinearity remained untouched. From the outset, let us remark that the
power (i.e. $k$) in the power - law Maxwell theory can not be arbitrary but
has to satisfy (at least) some of the energy conditions which are discussed
in the Appendix. It is demonstrated that pure magnetically charged black
holes do not exist even in this formalism. It is known that the interest in
NED aroused long ago during 1930's with the hopes to eliminate divergences
due to point charges. However, it is proved in this paper that according to
the value of the power Maxwell parameter in connection with energy
condititions, the solutions admit regular and naked singular
characteristics. Occurrence of naked singularities is known to violate the
cosmic censorship hypothesis. Understanding and the resolution of naked
singularities in general relativity remain one of the most challenging
problems to be solved. It is widely believed that the scales where this
singularity forms, classical attempts toward the resolution should be
replaced by the quantum theory of gravity. This motivates us to investigate
the formation and stability of naked singularities within the framework of
quantum mechanics. Our analysis will be based on the criterion of Horowitz
and Marolf \cite{17}\ (HM) in which quantum test particles obey the
Klein-Gordon and the Dirac equations are used to probe a naked singularity.
The criterion of HM has been used in different spacetimes to investigate
such classically naked singular spacetimes i.e whether they remain singular
or not within the context of quantum mechanics \cite{18,19,20,21,22,23,24}.

Meanwhile, it must be admitted that the physical interpretation of the
magnetic solution, whether it is due to a magnetic monopole or a vortex,
remains unclear. Naturally, such interpretations become less clear in the
power-Maxwell case as opposed to the case of standard linear Maxwell theory.

The plan of the paper is as follows. In section II, the action of the
Einstein-power-Maxwell formalism, solutions to the field equations are
given. In section III, the occurrence of naked singularity is analysed
within the framework of quantum mechanics. First, the definition of quantum
singularities for general static spacetimes is reviewed and then the
Klein-Gordon and the Dirac fields are used to test the quantum singularity.
The paper ends with Conclusion in section IV.

\section{The Solution and Spacetime Structure}

We start with the $3-$dimensional action in Einstein-power-Maxwell theory of
gravity with a cosmological constant $\Lambda $ ($8\pi G=1$)

\begin{equation}
I=\frac{1}{2}\int dx^{3}\sqrt{-g}\left( R-\frac{2}{3}\Lambda -\mathcal{F}%
^{k}\right) ,
\end{equation}%
in which $\mathcal{F}$ is the magnetic Maxwell invariant defined by%
\begin{equation}
\mathcal{F}=F_{\mu \nu }F^{\mu \nu }.  \notag
\end{equation}%
The field $2-$form is given by 
\begin{equation}
\mathbf{F}=B\left( r\right) dr\wedge d\theta ,
\end{equation}%
where $B\left( r\right) $ stands for the magnetic field to be determined.
Our metric ansatz for $3-$dimensions, is chosen as 
\begin{equation}
ds^{2}=-f_{1}\left( r\right) dt^{2}+\frac{dr^{2}}{f_{2}\left( r\right) }%
+f_{3}\left( r\right) d\theta ^{2},
\end{equation}%
in which $f_{i}\left( r\right) $ are some unknown functions to be found. The
parameter $k$ in the action is a real constant which is restricted by the
energy conditions (see the Appendix). Note that $k=1$ is a linear Maxwell
limit and in our treatments we consider the case $k\neq 1,$ so that our
treatment do not cover the linear Maxwell limit. The variation with respect
to the gauge potential yields the Maxwell equation 
\begin{equation}
\mathbf{d}\left( ^{\star }\mathbf{F}\mathcal{F}^{k-1}\right) =0,
\end{equation}%
where $^{\star }$ means duality and $\mathbf{d}\left( .\right) $ stands for
the exterior derivative. Remaining field equations are

\begin{equation}
G_{\mu }^{\nu }+\frac{1}{3}\Lambda \delta _{\ \mu }^{\nu }=T_{\mu }^{\nu },
\end{equation}%
in which 
\begin{equation}
T_{\mu }^{\nu }=-\frac{1}{2}\left( \delta _{\ \mu }^{\nu }\mathcal{F}%
^{k}-4k\left( F_{\mu \lambda }F^{\ \nu \lambda }\right) \mathcal{F}%
^{k-1}\right) ,
\end{equation}%
is the energy-momentum tensor due to the NED. It is readily seen that for $%
k=1$ all the foregoing expressions reduce to those of the standard linear
Maxwell theory. Nonlinear Maxwell equation (4) determines the unknown
magnetic field in the form%
\begin{equation}
B^{2}=\frac{f_{3}\left( r\right) }{f_{2}\left( r\right) }\frac{P^{2}}{%
f_{1}\left( r\right) ^{\frac{1}{2k-1}}},
\end{equation}%
in which $P$ is interpreted as the magnetic charge. Imposing this into the
energy-momentum tensor (6) results in%
\begin{equation}
T_{\ \nu }^{\mu }=\frac{1}{2}\mathcal{F}^{k}\text{diag}\left(
-1,2k-1,2k-1\right) ,
\end{equation}%
and the explicit form of $\mathcal{F}$ is given by 
\begin{equation}
\mathcal{F}=2\frac{P^{2}}{f_{1}\left( r\right) ^{\frac{1}{2k-1}}}.
\end{equation}%
The exact solution comes after solving the Einstein equations (5), which is
expressed by the metric functions%
\begin{eqnarray}
f_{1}\left( r\right) &\equiv &A\left( r\right) =-M+\frac{\left\vert \Lambda
\right\vert }{3}r^{2}=\frac{\left\vert \Lambda \right\vert }{3}\left(
r^{2}-r_{+}^{2}\right) , \\
f_{2}\left( r\right) &=&\frac{1}{r^{2}}\left( r^{2}+\frac{9\tilde{P}%
^{2}\left( 2k-1\right) ^{2}}{\left( k-1\right) \Lambda ^{2}}A\left( r\right)
^{\frac{k-1}{2k-1}}\right) A\left( r\right) , \\
f_{3}\left( r\right) &=&\frac{r^{2}}{A\left( r\right) }f_{2}\left( r\right) ,%
\text{ \ \ \ \ }k\neq 1,
\end{eqnarray}%
where $M$ may be interpreted as the mass and $\tilde{P}^{2}=2^{k-1}P^{2k}.$
We note that $r_{+}^{2}=\left\vert \frac{3M}{\Lambda }\right\vert ,$ and it
should not be taken as a horizon radius since our solution does not
represent a black hole. One finds the Ricci and Kretschmann scalars as%
\begin{eqnarray}
R &=&-2\left\vert \Lambda \right\vert -8\tilde{P}^{2}\left( k-\frac{3}{4}%
\right) A^{-\frac{k}{2k-1}}, \\
\mathcal{K} &=&\frac{4}{3}\Lambda ^{2}+\frac{32}{3}\tilde{P}^{2}\left( k-%
\frac{3}{4}\right) \left\vert \Lambda \right\vert A^{-\frac{k}{2k-1}%
}+4\left( 8k\left( k-1\right) +3\right) \tilde{P}^{4}A^{-\frac{2k}{2k-1}}.
\end{eqnarray}%
As one observes, depending on $k,$ one can put the solution into three
general categories. In the first category, $\frac{1}{4}\leq k<\frac{1}{2},$
and therefore $R$ and $\mathcal{K}$ are regular as the WEC and SEC (see
Appendix) are both satisfied. Since$,$ we may have $f_{3}\left( r_{\circ
}\right) =0$ for some $r_{\circ }$ it suggests that our coordinate patch is
not complete and needs to be revised. In such case we set 
\begin{equation}
x^{2}=r^{2}-r_{\circ }^{2}
\end{equation}%
which leads to the line element 
\begin{equation}
ds^{2}=-g_{1}\left( x\right) dt^{2}+\frac{dx^{2}}{g_{2}\left( x\right) }%
+g_{3}\left( x\right) d\theta ^{2}
\end{equation}%
with the metric functions

\begin{eqnarray}
g_{1}\left( x\right) &=&\frac{\left\vert \Lambda \right\vert }{3}\left(
x^{2}+r_{\circ }^{2}-r_{+}^{2}\right) , \\
g_{2}\left( x\right) &=&\left( x^{2}+r_{\circ }^{2}-\frac{9\tilde{P}%
^{2}\left( 2k-1\right) ^{2}}{\left\vert k-1\right\vert \Lambda ^{2}}%
g_{1}\left( x\right) ^{\frac{k-1}{2k-1}}\right) \frac{g_{1}\left( x\right) }{%
x^{2}} \\
g_{3}\left( x\right) &=&\left( x^{2}+r_{\circ }^{2}-\frac{9\tilde{P}%
^{2}\left( 2k-1\right) ^{2}}{\left\vert k-1\right\vert \Lambda ^{2}}%
g_{1}\left( x\right) ^{\frac{k-1}{2k-1}}\right) ,\text{ }k\neq 1.
\end{eqnarray}%
Here, one can show that for $x\in \left[ 0,\infty \right) $ then $%
g_{3}\left( x\right) <0,$ which implies a non-physical solution and hence
the power in this interval $\frac{1}{4}\leq k<\frac{1}{2}$ should be
excluded. The second category of solutions can be found by setting $\frac{1}{%
2}<k<1$ in which $g_{3}\left( x\right) >0$ possessing a non-singular
solution. It should be noted that the case for $k=1$ is already considered
in \cite{2,3,4,5} and the resulting spacetime has no curvature singularity.
The third category of solutions is when $k>1$ which results in a curvature
singularity. Therefore, by shifting the coordinate in accordance with $%
y^{2}=r^{2}-r_{+}^{2}$ we relocate the singularity to the point $y=0$ which
will be a naked singularity and our interest in this paper will be confined
entirely to this third category of solutions. In this new coordinate the
line element reads as%
\begin{equation}
ds^{2}=-h_{1}\left( y\right) dt^{2}+\frac{dy^{2}}{h_{2}\left( y\right) }%
+h_{3}\left( y\right) d\theta ^{2}
\end{equation}%
\begin{eqnarray}
h_{1}\left( y\right) &=&\frac{1}{3}\left\vert \Lambda \right\vert y^{2}, \\
h_{2}\left( y\right) &=&\left( y^{2}+r_{+}^{2}+\frac{9\tilde{P}^{2}\left(
2k-1\right) ^{2}}{\left( k-1\right) \Lambda ^{2}}\left( \frac{1}{3}%
\left\vert \Lambda \right\vert y^{2}\right) ^{\frac{k-1}{2k-1}}\right)
\left( \frac{\left\vert \Lambda \right\vert }{3}\right) \\
h_{3}\left( y\right) &=&\frac{3}{\left\vert \Lambda \right\vert }h_{2}\left(
y\right) ,\text{ \ \ \ \ \ \ }k\neq 1.
\end{eqnarray}%
with the scalars 
\begin{eqnarray}
R &=&-2\left\vert \Lambda \right\vert -8\tilde{P}^{2}\left( k-\frac{3}{4}%
\right) \left( \frac{1}{3}\left\vert \Lambda \right\vert y^{2}\right) ^{-%
\frac{k}{2k-1}}, \\
\mathcal{K} &=&\frac{4}{3}\Lambda ^{2}+\frac{32}{3}\tilde{P}^{2}\left( k-%
\frac{3}{4}\right) \left\vert \Lambda \right\vert \left( \frac{1}{3}%
\left\vert \Lambda \right\vert y^{2}\right) ^{-\frac{k}{2k-1}}+4\left(
8k\left( k-1\right) +3\right) \tilde{P}^{4}\left( \frac{1}{3}\left\vert
\Lambda \right\vert y^{2}\right) ^{-\frac{2k}{2k-1}}.
\end{eqnarray}

It can be seen that for $k>1,$ both $R$ and $\mathcal{K}$ are singular at $%
y=0$, and this singularity can easily be shown to be timelike.

Finally we add here that in the same frame but with an electric field
matter, there exist a black hole solution whose physical properties is
considered in a separate study \cite{25}.

\section{Singularity Analysis}

It has been emphasized in section II that the solution admits classical
naked singularity if the parameter $k>1$. This property is in fact one of
the most important consequences of the power - Maxwell field. Because the
previously obtained magnetically charged solution in $2+1$ dimensional
geometry with $k=1$ is regular \cite{2,3,4,5}. Naked singularities are one
of the "unlikable" predictions of the classical general relativity. The
reason is the cosmic censorship conjecture which forbids the formation of
classical naked singularities. Therefore, the resolution of these
singularities stand as an extremely important problem to be solved. Since
naked singularity occurs at very small scales where classical general
relativity is expected to be replaced by quantum theory of gravity, it is
worth to investigate the nature of this singularity with quantum test
fields. In probing the singularity, quantum test particles/fields obeying
the Klein-Gordon and Dirac equations are used. Our analysis will be based on
the pioneering work of Wald \cite{26}, which was further developed by
Horowitz and Marolf (HM) to probe the classical singularities with quantum
test particles obeying the Klein-Gordon equation in static spacetimes having
timelike singularities. According to HM, the singular character of the
spacetime is defined as the ambiguity in the evolution of the wave
functions. That is to say, the singular character is determined in terms of
the ambiguity when attempting to find self-adjoint extension of the operator
to the entire Hilbert space. If the extension is unique, it is said that the
space is quantum mechanically regular. The brief review is as follows:

\subsection{Quantum Singularities}

Consider a static spacetime $\left( M,g_{\mu \nu }\right) $\ with a timelike
Killing vector field $\xi ^{\mu }$. Let $t$ denote the Killing parameter and 
$\Sigma $\ denote a static slice.The Klein-Gordon equation in this space is

\begin{equation}
\left( \nabla ^{\mu }\nabla _{\mu }-M^{2}\right) \psi =0.
\end{equation}%
This equation can be written in the form

\begin{equation}
\frac{\partial ^{2}\psi }{\partial t^{2}}=\sqrt{f}D^{i}\left( \sqrt{f}%
D_{i}\psi \right) -fM^{2}\psi =-A\psi ,
\end{equation}%
in which $f=-\xi ^{\mu }\xi _{\mu }$ and $D_{i}$ is the spatial covariant
derivative on $\Sigma $. The Hilbert space $\mathcal{H}$, $\left(
L^{2}\left( \Sigma \right) \right) $\ is the space of square integrable
functions on $\Sigma $. The domain of the operator $A,$ $D(A)$ is taken in
such a way that it does not enclose the spacetime singularities. An
appropriate set is $C_{0}^{\infty }\left( \Sigma \right) $, the set of
smooth functions with compact support on $\Sigma $. Operator $A$ is real,
positive and symmetric therefore its self-adjoint extensions always exist.
If \ it has a unique extension $A_{E},$ then $A$ is called essentially
self-adjoint \cite{27,28,29}. Accordingly, the Klein-Gordon equation for a
free particle satisfies

\begin{equation}
i\frac{d\psi}{dt}=\sqrt{A_{E}}\psi,
\end{equation}
with the solution

\begin{equation}
\psi \left( t\right) =\exp \left[ -it\sqrt{A_{E}}\right] \psi \left(
0\right) .
\end{equation}%
If $A$ is not essentially self-adjoint, the future time evolution of the
wave function (29) is ambiguous. Then, HM criterion defines the spacetime
quantum mechanically singular. However, if there is only a single
self-adjoint extension, the operator $A$ is said to be\ essentially
self-adjoint and the quantum evolution described by Eq.(29) is uniquely
determined by the initial conditions. According to the HM criterion, this
spacetime is said to be quantum mechanically non-singular. In order to
determine the number of self-adjoint extensions, the concept of deficiency
indices is used. The deficiency subspaces $N_{\pm }$ are defined by ( see
Ref. \cite{30}for a detailed mathematical background),

\begin{align}
N_{+}& =\{\psi \in D(A^{\ast }),\text{ \ \ \ \ \ \ }A^{\ast }\psi =Z_{+}\psi
,\text{ \ \ \ \ \ }ImZ_{+}>0\}\text{ \ \ \ \ \ with dimension }n_{+} \\
N_{-}& =\{\psi \in D(A^{\ast }),\text{ \ \ \ \ \ \ }A^{\ast }\psi =Z_{-}\psi
,\text{ \ \ \ \ \ }ImZ_{-}<0\}\text{ \ \ \ \ \ with dimension }n_{-}  \notag
\end{align}%
The dimensions $\left( \text{ }n_{+},n_{-}\right) $ are the deficiency
indices of the operator $A$. The indices $n_{+}(n_{-})$ are completely
independent of the choice of $Z_{+}(Z_{-})$ depending only on whether $Z$
lies in the upper (lower) half complex plane. Generally one takes $%
Z_{+}=i\lambda $ and $Z_{-}=-i\lambda $ , where $\lambda $ is an arbitrary
positive constant necessary for dimensional reasons. The determination of
deficiency indices then reduces to counting the number of solutions of $%
A^{\ast }\psi =Z\psi $ ; (for $\lambda =1$),

\begin{equation}
A^{\ast }\psi \pm i\psi =0
\end{equation}%
that belong to the Hilbert space $\mathcal{H}$. If there is no square
integrable solutions ( i.e. $n_{+}=n_{-}=0)$, the operator $A$ possesses a
unique self-adjoint extension and it is essentially self-adjoint.
Consequently, a sufficient condition for the operator $A$ to be essentially
self-adjoint is to investigate the solutions satisfying Eq. (31) that do not
belong to the Hilbert space.

\subsection{Klein-Gordon Fields}

It was previously stated that the obtained solution is naked singular for $%
k>1.$ Quantum singularity analysis is almost hopeless for technical reasons
if the analysis is for any $k>1.$ Therefore, we restrict our analysis to a
specific parameter $k=2.$ This specific choice simplifies the metric which
is given by,

\begin{equation}
ds^{2}=-h_{1}\left( y\right) dt^{2}+\frac{dy^{2}}{\tilde{h}_{2}\left(
y\right) }+\tilde{h}_{3}\left( y\right) d\theta ^{2}
\end{equation}

\ 
\begin{eqnarray}
h_{1}\left( y\right) &=&\frac{1}{3}\left\vert \Lambda \right\vert y^{2}, \\
\tilde{h}_{2}\left( y\right) &=&\left( y^{2}+r_{+}^{2}+\alpha y^{2/3}\right) 
\frac{\left\vert \Lambda \right\vert }{3}, \\
\tilde{h}_{3}\left( y\right) &=&\frac{3}{\left\vert \Lambda \right\vert }%
\tilde{h}_{2}\left( y\right) ,\text{ }
\end{eqnarray}%
where $\alpha =\frac{81\tilde{P}^{2}}{\sqrt[3]{3}\left\vert \Lambda
\right\vert ^{5/3}}>0$ is a constant. The Kretschmann scalar for this
particular, $k=2$ is given by

\begin{equation}
\mathcal{K}=\frac{4}{3}\Lambda ^{2}-\frac{40\tilde{P}^{2}\left\vert \Lambda
\right\vert ^{1/3}}{\sqrt[3]{3}y^{4/3}}+\frac{\left( 76\tilde{P}^{4}\right)
3^{4/3}}{\left\vert \Lambda \right\vert ^{4/3}y^{8/3}}.
\end{equation}%
Clearly $y=0$ is a true curvature singularity. Upon separation of variables, 
$\psi =F(y)e^{in\theta },$ we obtain the radial portion of Eq.(31) as

\begin{equation}
\frac{d^{2}F\left( y\right) }{dy^{2}}+\frac{1}{y}\left\{ 1+\frac{y}{\tilde{h}%
_{2}\left( y\right) }\frac{d\left( \tilde{h}_{2}\left( y\right) \right) }{dy}%
\right\} \frac{dF\left( y\right) }{dy}+\frac{1}{\tilde{h}_{2}\left( y\right) 
}\left\{ \frac{c}{\tilde{h}_{3}\left( y\right) }-M\pm \frac{i}{h_{1}\left(
y\right) }\right\} F\left( y\right) =0
\end{equation}%
where $c\in 
\mathbb{R}
$ is a separation constant. Since the singularity is at $y=0,$ for small
values of $y$ each term in the above equation simplifies for massless $(M=0)$
case to

\begin{equation}
\frac{d^{2}F\left( y\right) }{dy^{2}}+\frac{1}{y}\frac{dF\left( y\right) }{dy%
}\pm \frac{\nu ^{2}}{y^{2}}iF\left( y\right) =0,
\end{equation}%
where $\nu ^{2}=\frac{9}{\left\vert \Lambda \right\vert ^{2}r_{+}^{2}}>0,$
whose solution is

\begin{equation}
F(y)=C_{1\nu }y^{\sqrt{\pm i}\nu }+C_{2\nu }y^{-\sqrt{\pm i}\nu },
\end{equation}%
in which $C_{1\nu }$ and $C_{2\nu }$ are arbitrary constants. In order to
check the square integrability, we define the function space on each $t=$%
constant hypersurface $\Sigma $ as $\mathcal{H}=\{F\mid \Vert F\Vert <\infty
\}$ with the following norm given for the metric (32) as,

\begin{equation}
\Vert F\Vert ^{2}=\frac{q^{2}}{2}\int_{0}^{\text{constant}}\frac{1}{\sqrt{%
h_{1}\left( y\right) }}\sqrt{\frac{\tilde{h}_{3}\left( y\right) }{\tilde{h}%
_{2}\left( y\right) }}\left\vert F\right\vert ^{2}dy\sim \int_{0}^{\text{%
constant}}\frac{\left\vert F\right\vert ^{2}}{y}dy,
\end{equation}%
where $q$ is a constant parameter. The above solution is checked for the
square integrablity near $y=0,$ for each sign of the solution found in Eq.
(39). The solution is square integrable if and only if the constant
parameter $C_{2n}=0,$ such that for each sign of Eq.(39) we have,

\begin{equation}
\Vert F\Vert ^{2}\sim \int_{0}^{\text{constant}}y^{\sqrt{2}\nu -1}dy=\frac{%
y^{\sqrt{2}\nu }}{\sqrt{2}\nu }\mid _{0}^{\text{constant}}<\infty .
\end{equation}
Therefore the operator $A$ has deficiency indices $n_{+}=n_{-}=1$, and is
not essentially self-adjoint, so that the spacetime is quantum-mechanically
singular.

\subsection{Dirac Fields}

The Dirac equation in $3D$ curved spacetime for a free particle with mass $m$
is given by,

\begin{equation}
i\sigma ^{\mu }\left( x\right) \left[ \partial _{\mu }-\Gamma _{\mu }\left(
x\right) \right] \Psi \left( x\right) =m\Psi \left( x\right) ,
\end{equation}%
where $\Gamma _{\mu }\left( x\right) $\ is the spinorial affine connection
given by

\begin{equation}
\Gamma_{\mu}\left( x\right) =\frac{1}{4}g_{\lambda\alpha}\left[ e_{\nu,\mu
}^{\left( i\right) }(x)e_{\left( i\right) }^{\alpha}(x)-\Gamma_{\nu\mu
}^{\alpha}\left( x\right) \right] s^{\lambda\nu}(x),
\end{equation}

\begin{equation}
s^{\lambda \nu }(x)=\frac{1}{2}\left[ \sigma ^{\lambda }\left( x\right)
,\sigma ^{\nu }\left( x\right) \right] .
\end{equation}

Since the fermions have only one spin polarization in $3D$ \cite{31}, the
Dirac matrices $\gamma ^{\left( j\right) }$ can be given in terms of Pauli
spin matrices $\sigma ^{\left( i\right) }$ \cite{32} so that

\begin{equation}
\gamma ^{\left( j\right) }=\left( \sigma ^{\left( 3\right) },i\sigma
^{\left( 1\right) },i\sigma ^{\left( 2\right) }\right) ,
\end{equation}%
where the Latin indices represent internal (local) frame. In this way,

\begin{equation}
\left\{ \gamma ^{\left( i\right) },\gamma ^{\left( j\right) }\right\} =2\eta
^{\left( ij\right) }I_{2\times 2},
\end{equation}%
where $\eta ^{\left( ij\right) }$\ is the Minkowski metric in $3D$ and $%
I_{2\times 2}$\ is the identity matrix. The coordinate dependent metric
tensor $g_{\mu \nu }\left( x\right) $\ and matrices $\sigma ^{\mu }\left(
x\right) $\ are related to the triads $e_{\mu }^{\left( i\right) }\left(
x\right) $\ by

\begin{align}
g_{\mu \nu }\left( x\right) & =e_{\mu }^{\left( i\right) }\left( x\right)
e_{\nu }^{\left( j\right) }\left( x\right) \eta _{\left( ij\right) }, \\
\sigma ^{\mu }\left( x\right) & =e_{\left( i\right) }^{\mu }\gamma ^{\left(
i\right) },  \notag
\end{align}%
where $\mu $\ and $\nu $\ stand for the external (global) indices. The
suitable triads for the metric (32) are given by,

\begin{equation}
e_{\mu }^{\left( i\right) }\left( t,y,\theta \right) =diag\left( y\sqrt{%
\frac{\left\vert \Lambda \right\vert }{3}},\left( \frac{3}{\left\vert
\Lambda \right\vert \left( y^{2}+r_{+}^{2}+\alpha y^{2/3}\right) }\right) ^{%
\frac{1}{2}},\left( y^{2}+r_{+}^{2}+\alpha y^{2/3}\right) ^{1/2}\right) ,
\end{equation}

The coordinate dependent gamma matrices and the spinorial affine connection
are given by

\begin{align}
\sigma ^{\mu }\left( x\right) & =\left( \left( \sqrt{\frac{3}{\left\vert
\Lambda \right\vert }}\right) \frac{\sigma ^{\left( 3\right) }}{y},i\left( 
\frac{\left\vert \Lambda \right\vert \left( y^{2}+r_{+}^{2}+\alpha
y^{2/3}\right) }{3}\right) ^{\frac{1}{2}}\sigma ^{\left( 1\right) },\frac{%
i\sigma ^{\left( 2\right) }}{\left( y^{2}+r_{+}^{2}+\alpha y^{2/3}\right)
^{1/2}}\right) , \\
\Gamma _{\mu }\left( x\right) & =\left( \frac{\left\vert \Lambda \right\vert
\left( y^{2}+r_{+}^{2}+\alpha y^{2/3}\right) ^{\frac{1}{2}}\sigma ^{\left(
2\right) }}{6},0,\frac{i\sqrt{\left\vert \Lambda \right\vert }}{6y^{1/3}%
\sqrt{3}}\left( 3y^{4/3}+\alpha \right) \sigma ^{\left( 3\right) }\right) . 
\notag
\end{align}%
Now, for the spinor

\begin{equation}
\Psi =\left( 
\begin{array}{c}
\psi _{1} \\ 
\psi _{2}%
\end{array}%
\right) ,
\end{equation}%
the Dirac equation can be written as

\bigskip 
\begin{eqnarray}
&&\frac{i}{y}\sqrt{\frac{3}{\left\vert \Lambda \right\vert }}\frac{\partial
\psi _{1}}{\partial t}-\left( \frac{\left\vert \Lambda \right\vert \left(
y^{2}+r_{+}^{2}+\alpha y^{2/3}\right) }{3})\right) ^{\frac{1}{2}}\frac{%
\partial \psi _{2}}{\partial y}+\frac{i}{\sqrt{\left( y^{2}+r_{+}^{2}+\alpha
y^{2/3}\right) }}\frac{\partial \psi _{2}}{\partial \theta } \\
&&-\left( \frac{\sqrt{\left\vert \Lambda \right\vert }\left( 3y^{4/3}+\alpha
\right) }{6y^{1/3}\sqrt{3}\left( y^{2}+r_{+}^{2}+\alpha y^{2/3}\right) ^{1/2}%
}+\frac{\sqrt{3\left\vert \Lambda \right\vert \left( y^{2}+r_{+}^{2}+\alpha
y^{2/3}\right) }}{6y}\right) \psi _{2}-m\psi _{1}=0,  \notag
\end{eqnarray}

\begin{eqnarray}
&&-\frac{i}{y}\sqrt{\frac{3}{\left\vert \Lambda \right\vert }}\frac{\partial
\psi _{2}}{\partial t}-\left( \frac{\left\vert \Lambda \right\vert \left(
y^{2}+r_{+}^{2}+\alpha y^{2/3}\right) }{3})\right) ^{\frac{1}{2}}\frac{%
\partial \psi _{1}}{\partial y}-\frac{i}{\sqrt{\left( y^{2}+r_{+}^{2}+\alpha
y^{2/3}\right) }}\frac{\partial \psi _{1}}{\partial \theta } \\
&&-\left( \frac{\sqrt{\left\vert \Lambda \right\vert }\left( 3y^{4/3}+\alpha
\right) }{6y^{1/3}\sqrt{3}\left( y^{2}+r_{+}^{2}+\alpha y^{2/3}\right) ^{1/2}%
}+\frac{\sqrt{3\left\vert \Lambda \right\vert \left( y^{2}+r_{+}^{2}+\alpha
y^{2/3}\right) }}{6y}\right) \psi _{1}-m\psi _{2}=0  \notag
\end{eqnarray}

\bigskip The following ansatz will be employed for the positive frequency
solutions:

\begin{equation}
\Psi _{n,E}\left( t,x\right) =\left( 
\begin{array}{c}
Z_{1n}(y) \\ 
Z_{2n}(y)e^{i\theta }%
\end{array}%
\right) e^{in\theta }e^{-iEt}.
\end{equation}

The radial part of the Dirac equation becomes,

\begin{eqnarray}
&&Z_{2n}^{^{\prime }}(y)+\left\{ \frac{\sqrt{3}\left( n+1\right) }{\sqrt{%
\left\vert \Lambda \right\vert }\left( y^{2}+r_{+}^{2}+\alpha y^{2/3}\right) 
}+\frac{\left( 3y^{4/3}+\alpha \right) }{6y^{1/3}\left(
y^{2}+r_{+}^{2}+\alpha y^{2/3}\right) }+\frac{1}{2y}\right\} Z_{2n}(y)+ \\
&&\frac{1}{\sqrt{\left( y^{2}+r_{+}^{2}+\alpha y^{2/3}\right) }}\left\{ m%
\sqrt{\frac{3}{\left\vert \Lambda \right\vert }}-\frac{3E}{\left\vert
\Lambda \right\vert y}\right\} Z_{1n}(y)e^{-i\theta }=0  \notag
\end{eqnarray}

\begin{eqnarray}
&&Z_{1n}^{^{\prime }}(y)+\left\{ -\frac{\sqrt{3}n}{\sqrt{\left\vert \Lambda
\right\vert }\left( y^{2}+r_{+}^{2}+\alpha y^{2/3}\right) }+\frac{\left(
3y^{4/3}+\alpha \right) }{6y^{1/3}\left( y^{2}+r_{+}^{2}+\alpha
y^{2/3}\right) }+\frac{1}{2y}\right\} Z_{1n}(y)+ \\
&&\frac{1}{\sqrt{\left( y^{2}+r_{+}^{2}+\alpha y^{2/3}\right) }}\left\{ m%
\sqrt{\frac{3}{\left\vert \Lambda \right\vert }}+\frac{3E}{\left\vert
\Lambda \right\vert y}\right\} Z_{2n}(y)e^{i\theta }=0  \notag
\end{eqnarray}

The behavior of the Dirac equation near $y=0$ reduces to,

\begin{equation}
Z_{j}^{^{\prime \prime }}(y)+\frac{2}{y}Z_{j}^{^{\prime }}(y)+\frac{\beta
^{2}}{y^{2}}Z_{j}(y)=0,\text{ \ \ \ \ \ \ \ \ }j=1,2
\end{equation}%
where $\beta ^{2}=\frac{1}{4}+\left( \frac{3E}{\left\vert \Lambda
\right\vert r_{+}}\right) ^{2}.$ The solution is given by

\begin{equation}
Z_{j}(y)=C_{1j}y^{\gamma _{1}}+C_{2j}y^{\gamma _{2}},
\end{equation}%
where $C_{1j}$ and $C_{2j}$ are arbitrary constants and exponents are given
by

\begin{equation}
\gamma _{1}=-\frac{1}{2}+i\frac{3\left\vert E\right\vert }{\left\vert
\Lambda \right\vert r_{+}},\text{ \ \ \ \ \ \ \ \ \ }\gamma _{2}=-\frac{1}{2}%
-i\frac{3\left\vert E\right\vert }{\left\vert \Lambda \right\vert r_{+}}%
\text{.}  \notag
\end{equation}%
The condition for the Dirac operator to be quantum - mechanically regular
requires that both solutions should belong to the Hilbert space $\mathcal{H}$%
. Squared norm for this solution

\begin{equation}
\sim \int_{0}^{\text{constant}}\frac{\left\vert Z_{j}(y)\right\vert ^{2}}{y}%
dy\sim \int_{0}^{\text{constant}}y^{-2}dy\sim \frac{1}{y}\mid _{0}^{\text{%
constant}}\rightarrow \infty ,\text{ \ \ \ \ \ }
\end{equation}%
diverges. This implies that solution do not belong to the Hilbert space.
Consequently, if the classical singularity at $y=0$ is probed with fermions
the spacetime behaves quantum mechanically singular.

\section{Conclusion}

In this paper, a new class of magnetically charged solutions in $3D$
Einstein-Power-Maxwell theory has been presented. As in the linear Maxwell
case, our solutions do not admit black holes but apart from the linear
Maxwell case the power-law Maxwell theory admits singular solutions as well.
The main contribution of the nonlinear Maxwell field in our solutions is to
create timelike naked singularities for specific values of parameter $k>1$
which is non-existent in the linear theory. \ This singularity has been
analysed from the quantum mechanical point of view. Quantum test particles
obeying the Klein-Gordon and the Dirac equations are used to probe the
singularity.

The analysis of the naked singularity from quantum mechanical point of view
has revealed that the considered spacetime is generically quantum singular
when it is probed with fields obeying Klein-Gordon and Dirac equations. It
is interesting to note that, in contrast to the considered spacetime, the
probe of naked singularity with Dirac fields in other $3D$ metrics, namely
BTZ \cite{20} and matter coupled BTZ \cite{23} spacetimes was shown to be
quantum mechanically regular. It is also shown in this study that for
general modes of spin zero Klein-Gordon fields, the spacetime is still
singular.

\textbf{Acknowledgment: }\textit{We wish to thank the anonymous referee for
his constructive and valuable suggestions.}

\textbf{APPENDIX: Energy Conditions}

When a matter field couples to any system, energy conditions must be
satisfied for physically acceptable solutions. We follow the steps as given
in \cite{33,34} to find the bounds of the power parameter $k$ of the Maxwell
field.

\subsection{Weak Energy Condition (WEC)}

The WEC states that,

\begin{equation}
\rho \geq 0\text{ \ \ \ \ \ \ \ \ \ \ \ and \ \ \ \ \ \ \ \ }\rho +p_{i}\geq
0\text{ \ \ \ \ \ }(i=1,2)  \tag{A1}
\end{equation}%
in which $\rho $ is the energy density and $p_{i}$ are the principal
pressures given by

\begin{equation}
\rho =-T_{t}^{t}=\frac{1}{2}\mathcal{F}^{k},\text{ \ \ \ \ \ \ \ \ }%
p_{i}=T_{i}^{i}=\frac{2k-1}{2}\mathcal{F}^{k}\text{ \ \ (no sum).}  \tag{A2}
\end{equation}%
This condition imposes that $k>0$.

\subsection{Strong Energy Condition (SEC)}

This condition states that;

\begin{equation}
\rho +\dsum\limits_{i=1}^{2}p_{i}\geq 0\text{ \ \ \ \ \ \ and \ \ \ \ \ \ }%
\rho +p_{i}\geq 0,  \tag{A3}
\end{equation}%
which amounts, together with the WEC to constrain the parameter $k$ $\geq 
\frac{1}{4}$.

\subsection{Dominant Energy Condition (DEC)}

In accordance with DEC, the effective pressure $p_{eff}$ should not be
negative i.e. $p_{eff}\geq0$ where

\begin{equation}
p_{eff}=\frac{1}{2}\dsum\limits_{i=1}^{2}T_{i}^{i}.  \tag{A4}
\end{equation}%
One can show that DEC, together with SEC and WEC impose the following
condition on the parameter $k$ as 
\begin{equation}
k>\frac{1}{2}.  \tag{A5}
\end{equation}

\subsection{Causality Condition (CC)}

In addition to the energy conditions one may impose the causality condition
(CC)

\begin{equation}
0\leq \frac{p_{eff}}{\rho }<1  \tag{A6}
\end{equation}%
which implies that

\begin{equation}
\frac{1}{2}\leq k<1.  \tag{A7}
\end{equation}%
The CC is clearly violated in our solutions since we abide by the parameter $%
k>1$, throughout the paper.

\bigskip

\end{document}